\documentclass[
twocolumn,
longbibliography,
showpacs,                     
showkeys,                  
secnumarabic,
amssymb, 
nobibnotes, 
floatfix,
aps, 
pre]{revtex4-1}

\usepackage{graphicx}
\usepackage{latexsym}
\usepackage{amsmath} 
\usepackage{verbatim}

\begin{document}

\title{The definition of the thermodynamic entropy
in statistical mechanics}

\author{Robert H. Swendsen}
\email[]{swendsen@cmu.edu}

\affiliation{The Department of Physics, Carnegie Mellon University,
Pittsburgh, PA 15213, USA}


\date{\today}

\begin{abstract}
A definition of the thermodynamic entropy
based on the time-dependent probability distribution 
of the macroscopic variables
is developed.
When a constraint
in a composite system
 is released,
the probability distribution for the new equilibrium values
goes to a narrow peak.
Defining the entropy by the logarithm of the probability distribution
automatically makes it a maximum at the equilibrium values,
so it satisfies  the Second Law.
It is also  satisfies
the 
postulates of thermodynamics.
Objections to this definition by
Dieks and Peters
are discussed 
and resolved.
\end{abstract}

\maketitle

\section{Introduction}

Thermodynamics is
an extremely successful phenomenological theory 
of macroscopic experiments.
The entropy plays a central role in this theory
because it is a unique function
for each system
 that determines
all thermodynamic information.
The calculation of the form of the entropy
lies in 
the microscopic description given by
statistical mechanics.
In this paper,
I present a simple derivation of the entropy 
using  reasonable assumptions about 
the probability distributions 
of macroscopic variables
and approximations based on
the large number of particles
in macroscopic systems.

The basic task
of  thermodynamics 
is the prediction of 
the values of 
 the macroscopic variables 
after the release of one or more constraints
and the subsequent relaxation to
a new equilibrium.
This appears in the key thermodynamic  postulate
that is a particular form of the second law.\cite{Tisza,Callen,RHS_book}.
\begin{quote}
The values assumed by the extensive parameters
of an isolated composite system in the absence of an internal
constraint are those that maximize the entropy over the set of
all constrained macroscopic states\cite{RHS_book}.
\end{quote}
I will show that the solution to this problem 
in statistical mechanics 
leads to a function that satisfies this postulate,
as well as satisfying the rest of the postulates of thermodynamics.
Since these postulates are sufficient to generate all
of thermodynamics,
and since the thermodynamic entropy is unique\cite{Lieb_Yngvason},
 this function can be identified as the entropy.

I have presented other derivations in the past that are equivalent,
though perhaps not as 
direct\cite{RHS_1,RHS_4,RHS_5,RHS_8,RHS_9,RHS_unnormalized}.
They have been  criticized
by  
Dieks\cite{Dieks_unique_entropy,Dieks_logic_of_identity_2014}
  and 
  Peters\cite{Peters_2010,Peters_2014},
  whose arguments  
  will be discussed  
  in Sections \ref{section: Dieks}
    and 
  \ref{section: Peters}.
\section{The prediction of equilibrium values from statistical mechanics}
\label{section: statistical mechanics}
 
Thermodynamics is 
a description of the properties of 
systems containing many particles
(macroscopic systems),
for which  the fluctuations 
can be ignored because they
are smaller than
the experimental resolution.
The basic problem 
of thermodynamics 
is to predict the equilibrium
values of the extensive variables
 after the release of a constraint in a composite system.
 I will first consider this as a problem is statistical mechanics,
 without using any thermodynamic concepts.

 Consider a composite system of 
 $M \ge 2$ subsystems,
 with a total energy 
 $E_T$,
 volume
 $V_T$,
 and 
 particle number 
 $N_T$\cite{footnote_1_particle_types}.
 Denote the total phase space for this composite system 
(in three dimensions)
by 
$\{p,q\}$,
where 
$p$ is the $3N_T$-dimensional momentum space,
and 
$q$ is the $3N_T$-dimensional configuration space.
Define the probability distribution 
in the phase space of  the composite system
as 
$\phi_T \left(   \{p, q\} , t  \right)$,
where $t$ is the time.
I'll assume that 
the composite system is initially
in equilibrium at time $t=0$,
and that
the initial conditions are given 
by setting 
$\phi_T$ 
equal to a constant,
subject to all information available about the system
at that time.

Assume that interactions between subsystems are sufficiently short-ranged
that they may be neglected\cite{RHS_continuous}.
Then, we can write the total Hamiltonian as 
a sum of contributions 
from each system.
\begin{equation}\label{H total 1}
H_T
=
\sum_{j=1}^M
H_j (E_j, V_j, N_j)
\end{equation}
The energy, volume, and particle number 
of subsystem $j$
 are  denoted as
$E_j$, $V_j$, and $N_j$,
subject to the conditions
on the sums.
\begin{equation}
\sum_{j=1}^{M} E_j   = E_T  ; \,
\sum_{j=1}^{M} V_j   = V_T  ; \,
\sum_{j=1}^{M} N_j   = N_T  
\end{equation}

In keeping with the idea that 
we are describing 
 macroscopic experiments,
 assume that 
 no measurements
are made that might identify 
individual particles,
whether or not they are formally  indistinguishable\cite{RHS_distinguishability}.
This means that
there are 
$N_T!/ \left( \prod_{j=1}^{M}  N_j!\right)$
different permutations for assigning particles to subsystems,
and all permutations may be regarded as equally probable.
 The probability distribution 
in the phase space of  the composite system
is given by
\begin{eqnarray}\label{phi t=0}
\phi_T \left(   \{p, q\} , t =0 \right)
&=&
\frac{1}{\Omega_T}
\left(  \frac{N_T! }{ \prod_{j=1}^{M}  N_j! }  \right) \nonumber  \\ 
&& \times
\prod_{k=1}^{M}
\delta \left( E_k - H_k( \{ p_k, q_k \} )  \right)   ,
\end{eqnarray}
 where 
$\{ p_k, q_k \}$ 
is the phase space for the particles in subsystem $k$,
and 
$\Omega_T$
is a normalization factor.
The constraint that 
the $N_k$ particles in
subsystem $k$
are restricted to a volume 
$V_k$
is left implicit in 
Eq.~(\ref{phi t=0}).

The probability distribution for the macroscopic observables 
can then be written as 
\begin{eqnarray}\label{W 1}
W \left( 
\{ E_j, V_j,N_j \}
 \right)
&=&
\frac{N_T!}{\Omega_T}
\left(
\frac{
1
}{
\prod_j N_j!
}
\right)  
  \nonumber  \\
&&
\times
\int  dp  \int   dq  
\prod_{j=1}^M
\delta( E_j - H_j )   ,
\end{eqnarray}
or
\begin{equation}\label{W 2}
W(
 \{E_j, V_j, N_j 
\}
  ) 
=
\frac{
\prod_{j=1}^M
\Omega_j ( E_j, V_j, N_j )
}{
\Omega_T / N_T! h^{3N_T}
}         ,
\end{equation}
where
\begin{equation}\label{Omega j 1}
\Omega_j
=
\frac{1}{h^{3N_j} N_j!}
\int_{-\infty}^{\infty}   dp_j  \int_{V_j}   dq_j   \,
\delta( E_j - H_j )   .
\end{equation}
The factor of $1/h^{3N_j}$,
where $h$ is Planck's constant,
is not necessary for classical mechanics.
It has been included  to ensure that the
final answer agrees 
with the classical  limit 
from  quantum statistical mechanics\cite{RHS_book}.

There is no requirement that the Hamiltonians
$H_j$
are the same,
so there is also no requirement 
that the individual $\Omega_j$'s
have the same functional form.
Long-range interactions 
\textit{within a system}
are  allowed.

If one or more constraints are now released,
the probability
$\phi_T \left(   \{p, q\} , t  \right)$
will become time dependent.
After sufficient time  has passed,
the probability distribution will have spread throughout 
the available phase space,
although it will still be non-uniform on the finest scale
due to Liouville's theorem.
The probability distribution for the macroscopic variables 
will again be given by
$W(
 \{E_j, V_j, N_j 
\})$,
but now without the constraints 
on the variables 
that have been 
released\cite{RHS_6}.
The functional dependence of $W$
on the variables
$\{E_j, V_j, N_j 
\}$
does not change when a constraint is released.

An important advantage of working with the probability distributions 
for  macroscopic observables  
is that they converge 
to the equilibrium probability  distributions 
at the end of an irreversible process\cite{RHS_6}.
Although it is not necessary,
the introduction of coarse graining\cite{Penrose_book}
or the modification of the microscopic 
probability distribution 
by invoking typicality\cite{Goldstein_Lebowitz,Lebowitz_symmetry}
 leads to the same results.

Usually,
$W$ is a very narrow function of the released variables.
The main exception is  the case of a first-order phase transition,
in which 
it can be a very broad function 
of the relevant variable\cite{RHS_continuous}.
This situation is discussed in Ref.~\cite{RHS_continuous},
and I will ignore it for the present discussion.

The location of the  narrow peak 
in $W$ as a function of the variable 
describing a released constraint
 gives the final equilibrium value of that variable
 at the end of the irreversible process.
For example,
if subsystems $1$ and $2$ 
are brought in thermal contact
so that energy transfer is possible,
the final value of $E_1$
would be given by the location of the 
maximum of $W$
to within thermal fluctuations.
This characterizes the equilibrium values as the mode 
of the probability distribution,
not the mean.
The difference between the mean and the mode 
is of the order of $1/N$,
which is very small and far less than 
the assumed experimental accuracy.
Indeed, it is not even measurable 
for macroscopic systems\cite{SW_2015_PR_E_R}.

When subsystems are separated,
the probability $W$ remains unchanged.
The constraint is restored, and 
the variable that was being exchanged
keeps its value,
which is known to within the very small fluctuations.

The normalization constant,
$\Omega_T$,
 is dependent on exactly 
which constraints might be released,
but the other factors are not.
Since the only property 
 of the function 
$W( \{E_j, V_j, N_j \})$
that is needed 
is that it has a very narrow
peak at the equilibrium value(s)
after the release of constraint(s),
the value of $\Omega_T$
does not affect the argument.

Now that   the probability distribution
for the equilibrium variables 
has been determined,
we can turn to the definition of entropy.

\section{The definition of the thermodynamic entropy}
\label{section: entropy}

Following Boltzmann\cite{Boltzmann,Boltzmann_translation,RHS_4},
the thermodynamic entropy 
may be identified as 
 the logarithm of the probability distribution
$W$, plus an arbitrary constant.
\begin{equation}\label{S = ln W 1}
S_T ( \{E_j, V_j, N_j\})
=
k_B \ln W 
+ X 
\end{equation}
Since the probability is a maximum at equilibrium,
 the entropy is also 
with this definition.
Although Boltzmann 
considered  a dimensionless entropy
and 
never used the 
``Boltzmann constant,''  $k_B$,
which was introduced by Planck\cite{Planck_1901,Planck_book},
I have included a factor of $k_B$
to be consistent with physical  units.

Combining  Eqs.~(\ref{W 2}), (\ref{Omega j 1})
and (\ref{S = ln W 1}),
 the total entropy can be written as a sum of 
 $M$ terms,
each of which depends only on the properties of a single subsystem,
plus a constant.
\begin{equation}\label{S = sum Omega j 1}
S_T   
=
\sum_{j=1}^M
S_j  ( E_j, V_j, N_j  )
-
k_B 
\ln \left[ \frac{ \Omega_T  }{ N_T! h^{3N_T} } \right]
+
X
\end{equation}
 The entropy of the $j$-th subsystem 
 in Eq.~(\ref{S = sum Omega j 1}) 
 is given by 
\begin{equation}\label{S j = kB ln Omega j}
S_j  ( E_j, V_j, N_j  )
=
k_B  \ln  \Omega_j ( E_j, V_j, N_j  )  ,
\end{equation}
or
\begin{equation}\label{S j = kB ln Omega j explicit}
S_j         
=
k_B  \ln 
\left[
\frac{1}{h^{3N_j} N_j!}
\int_{-\infty}^{\infty}   dp_j  \int_{V_j}   dq_j   \,
\delta( E_j - H_j )  
\right]  .
\end{equation}
The entropy of subsystem $j$
contains the factor $1/N_j!$,
which arises from the multinomial factor 
in Eq.~(\ref{phi t=0}).
It would be  possible to add
an arbitrary constant $X_j$ 
to $S_j$ in 
Eq.~(\ref{S j = kB ln Omega j explicit}),
but I have chosen to set
$X_j=0$
for all $j$, 
which is the usual  convention\cite{RHS_9}.

$S_j$  only depends on the properties of 
system $j$,
which means that the total entropy is separable.
This is just the usual thermodynamic property
of additivity,
but viewed from the perspective of
dividing up  a composite system,
rather than assembling one.

Since $\Omega_T$
has been defined to be a normalization constant,
if all chosen constraints are released,
the  value of $S_T$ 
after the composite system 
has returned to equilibrium
is given entirely by the additive constant
(neglecting terms of the order of the 
logarithm of the particle numbers).
\begin{equation}\label{S = X in equilibrium}
S_T   (\text{after release})
\rightarrow
X
\end{equation}
This will be true regardless of which 
constraints have been chosen  
to determine 
$\Omega_T$,
as long as all of those constraints are released.

A convenient choice of $X$ is 
$k_B \ln \left[  \Omega_T  / N_T! h^{3N_T} \right]$.
Then the total entropy of the composite system
is just given by the sum of the subsystem entropies.
But this choice is not required.

\section{The application of the entropy equations}

%
%

Eqs.~(\ref{S = sum Omega j 1}),
(\ref{S j = kB ln Omega j}),
 and  (\ref{S j = kB ln Omega j explicit})
are intended to be applied to the set of all systems in the world
that can be regarded as classical.
That includes not only systems in a particular laboratory,
but also those in a different city or continent.
Most systems will not interact with each other
because of physical separation,
and the constraints of their not exchanging
energy, volume, or particles
 are expected to remain indefinitely.

The entropy of a single system is given by Eq.~(\ref{S j = kB ln Omega j explicit}).
For experiments involving only a local  group of 
systems
(or subsystems of the overall composite system),
the existence of  many other (sub)systems
can be safely ignored, because 
their properties do not affect  
 the local thermodynamic variables.
Similarly,
the value of 
the additive constants in 
Eq.~(\ref{S = sum Omega j 1})
will not affect the predictions of any experiment.

%
%

Eqs.~(\ref{S = sum Omega j 1}) and (\ref{S j = kB ln Omega j explicit}) 
allow us to find  
the  non-negative
change in total entropy
($\Delta S_T \ge 0$)
during any 
irreversible process 
between equilibrium states
that occurs after the release of a constraint,
as well as 
the final 
equilibrium values  of thermodynamic observables.

Dieks has criticized this derivation of the entropy.
I discuss his views in the next section.

\section{Dieks'  objection}
\label{section: Dieks}

Dieks' criticism rests on the claim that the choice of additive constant, $X$
in Eq.~(\ref{S = sum Omega j 1}),
is essential for obtaining my results for the entropy\cite{Dieks_unique_entropy}.
This claim is untenable, 
 since I have derived 
the entropy of an arbitrary subsystem 
 [Eq.~(\ref{S j = kB ln Omega j explicit})]
without fixing the value of $X$,
and the value of $X$ has no physical consequences.

Looking further, we can see that
Dieks means something  different.
He is  interested in the value of the entropy 
of the entire composite system of $M$ subsystems
for the case in which all constraints have been released.
As  shown above in Eq.~(\ref{S = X in equilibrium}),
the release of all constraints leads to a constant
$S_T  
 \rightarrow X$,
where $X$ is arbitrary.
Dieks 
is concerned about the determination  
of a particular form of this constant.
Since there are no physical consequences 
for any value of $X$,
 I fail to see the importance of the issue.

Dieks explicitly recognizes that this issue is without importance.
Writing $N$ for what I have called $N_T$, he says
in a footnote:
\begin{quote}
A more detailed discussion should also take into account that the division by $N!$ is
without significance anyway as long as 
$N$ is constant\cite{Dieks_unique_entropy}.
\end{quote}
However,
he still uses the value of this constant to frame his objection 
to my definition.
The reason for this contradiction  
might lie in 
his incorrect description of my definition of entropy,
which he claims  amounts simply 
to dividing the traditional expression by
$N!$. 

I will consider his argument in detail.

\subsection{Two simple subsystems}

Dieks 
  considered
  an isolated composite system consisting 
of only  two ideal gases ($M=2$),
and
simplified his analysis by ignoring the 
energy dependence.
In discussing his  argument,
I will depart from Dieks' notation\cite{footnote_on_Dieks_notation} 
by using 
 $N_T=N_1+N_2$
 as the constant total number of particles 
to  be consistent with the notation I used in previous sections.
For clarity,
I will also retain an arbitrary value of the 
additive constant $X$
(see Eq.~(\ref{S = ln W 1}), above)
until the end of the discussion,
although Dieks 
makes the specific choice 
of
$X=k_B \ln \left( V_T^{N_T}/N_T! \right)$,
``for reasons of convenience,''
early 
in his argument\cite{Dieks_unique_entropy}.

For Dieks'
two subsystems of classical ideal gases, 
my 
Eq.~(\ref{S = sum Omega j 1})
 becomes
his Eq.~(2),
\begin{align}
S_T (N_1,V_1;N_2,V_2)
=&
k_B
\ln \left(
\frac{N_T!}{N_1! N_2!}
\frac{ V_1^{N_1}V_2^{N_2} }{ V_T^N}
\right)
+X           \nonumber         \\
=&
k_B
\ln \left(
\frac{V_1^{N_1}}{N_1! }
\right)
+
k_B
\ln \left(
\frac{V_2^{N_2}}{ N_2!}
\right)           \nonumber         \\
& 
 -
k_B
\ln \left(
\frac{ V_T^{N_T} }{ N_T!}
\right)
+X   ,
\label{Dieks 2}
\end{align}
where 
$N_T=N_1+N_2$
and
$V_T=V_1+V_2$
are constants.
Note that Dieks'
choice for the value of the constant $X$
means that the last two terms 
in Eq.~(\ref{Dieks 2})
cancel in his Eq.~(2).

Since Eq.~(\ref{Dieks 2})
is valid for all values of 
$N_1$, $N_2$, $V_1$, and $V_2$,
we immediately have the 
(partial) 
entropies, 
\begin{equation}\label{partial S j}
S_j (V_j,N_j )
=
k_B
\ln \left(
\frac{V_j^{N_j}}{ N_j!}
\right)         ,
\end{equation}
where $j=1$ or $2$.
I claim that this is a proper derivation of the factors 
$1/N_j!$.

Dieks made  the following  comment on 
his Eq.(2)   
(writing $N$ for what I have called  $N_T$).
\begin{quote}
Indeed, the dependence of the total entropy in 
Eq.~(2) on $N_1$ and $N_2$ is unrelated to
how $N$ occurs in this formula 
(and to the choice of the zero 
of the total entropy)\cite{Dieks_unique_entropy}.
\end{quote}
His comment   confirms  the 
validity of my derivation 
of the factors $1/N_1!$
and $1/N_2!$
in the entropies of subsystems 
$1$ and $2$,
as well as the irrelevance 
of the value 
of the additive constant 
$X$.

Dieks then calculates the entropy
\emph{after} the release of the constraint
on the particle number
and return to equilibrium.
He gets the result
$X=k_B \ln \left( V_T^{N_T}/N_T! \right)$.
Dieks   claims 
that 
this was the way I had obtained a
$-k_B \ln N_T!$
dependence of the total entropy.
I did not fix the value of $X$,
so I did not derive an expression 
for the entropy after the release of constraints.

Actually, 
 the form of the 
$k_B \ln \left( V_T^{N_T}/N_T! \right)$
term in the joint entropy does not come from choosing
 the constant $X$
to make 
$S_T = \sum_{j=1}^M S_j $,
but rather from the simplicity 
of the example used.
If the properties of the subsystems are
 generalized,
a different result is obtained.

\subsection{Two less simple subsystems}

Consider the  entropy,
\begin{equation}\label{S_with_energy_shift}
S_j
= 
k_B N_j
\left[
\frac{3}{2} 
\ln \left( \frac{E_j - N_j a_j  }{  N_j }  \right)
+
\ln \left( \frac{V_j  }{  N_j }  \right)
+
Y_j'
\right]   ,
\end{equation}
where I have used Stirling's approximation.
The total entropy before allowing the systems to interact is 
$S_T = S_1+S_2$.
The energy dependence is now given explicitly,
and an energy shift per particle, $a_j$, is given to each subsystem.
Assume that $a_1=0$ and $a_2>0$.
Let subsystems $1$ and $2$ come into thermal contact and exchange energy and particles.

The temperature dependence of the energy
in the $j$-th subsystem  is given by
\begin{equation}\label{U_j 1}
E_j = \frac{3}{2} k_B N_j  T_j + N_j a_j   ,
\end{equation}
so the condition of equilibrium with respect to energy exchange is
\begin{equation}
\frac{ E'_1 }{ N_1 }
=
\frac{ E'_2 }{ N_2} -  a_2    ,
\end{equation}
where I have indicated the new values of the energies by 
$E'_1$ and $E'_2$.

Now let the two subsystems exchange particles.
From the condition of equilibrium with respect to particle number,
it is straightforward to derive
\begin{equation}\label{n1_n2_T}
 \ln  \left(  \frac{ V_1  }{ N''_1  }  \right)
=   
 \ln  \left(  \frac{ V_2 }{ N''_2 }  \right)
-
\frac{3}{2} 
\left[
\frac{1}{E''_2/N''_2 a_2 - 1}
\right]      ,
\end{equation}
where I have indicated the new values of the energies and particle numbers 
by double primes, i.e: $E'_j$ and $N''_j$.
Since 
$E''_1/N''_1 \ne E''_2/N''_2$
and 
  $V_1/N''_1 \ne V_2/N''_2$,
  the total entropy
cannot be written as a function of 
$(E''_1+E''_2)$,
$(V_1+V_2)$,
and 
$(N''_1+N''_2)$.
There is no term in 
$S''_T = S''_1(E''_1,V_1,N''_1) + S''_2(E''_2 ,V_2,N''_2)$
of the form
$k_B \ln \left( V_T^{N_T}/N_T! \right)$.

For the next example it will be sufficient 
to again consider ideal gases and
ignore the energy dependence.

\subsection{Three simple subsystems}

Dieks' analysis 
does not recognize that
the thermodynamic variables in 
subsystems $1$ and $2$ 
remain $N_1$ and $N_2$,
even after the systems come to equilibrium.
They are not replaced by a single variable.
This can be seen most easily by considering 
$M \ge 3$ subsystems.
To avoid confusion,
denote the number of particles in
subsystems $1$ and $2$ by
 $N_{1,2}=N_1+N_2$,
 because it is no longer constant.
Now consider 
how subsystems $1$ and $2$ interact
with a third subsystem.

Let subsystems $1$ and $2$ 
first come to equilibrium
and then
be separated again,
denoting the new particle numbers by 
$N'_1$ and $N'_2$. 
Let subsystem $3$ 
originally have a high number density,
$N_3/V_3 > N'_1/V_1= N'_2/V_2$.
Now let subsystem $2$ exchange particles with 
system $3$,
so that $N_2$ increases 
($N''_2>N'_2$).
Subsystems $2$ and $3$ come to a new equilibrium,
for which  
\begin{equation}
\frac{N'_1}{V_1}
<
\frac{N''_2}{V_2}
=
\frac{N''_3}{V_3}     .
\end{equation}

The entropy of subsystems $1$ and $2$ is 
(with Stirling's approximation),
\begin{equation}
S'_1+S''_2
\approx
k_B
N'_1
\ln \left(
\frac{V_1}{ N'_1}
\right)    
+
k_B
N''_2
\ln \left(
\frac{V_2}{ N''_2}
\right)   .
\end{equation}
Since the number density is different in 
subsystems $1$ and $2$,
it is clear that
$S'_1+S''_2$
 is not given by 
 $k_B N_{1,2} \ln \left( V_{1,2} / N_{1,2} \right)$. 
\subsection{An arbitrary number of subsystems}

When Dieks discusses the case of many systems,
he writes that I require a ``consistency'' condition,
\begin{quote}
that the entropy formula should be such
that there will be no change in entropy
when a partition is removed\cite{Dieks_unique_entropy}.
\end{quote}
I do not require it, and
it is not  a consistency condition.
It is the condition that  systems 
separated by a partition
are in equilibrium,
which is not generally true 
in the presence of a constraint.

To summarize,
I have calculated the dependence of the entropy 
on the variables
$\{E_j, V_j, N_j  \vert =1, \dots,M\}$
in the presence or absence of arbitrary constraints.
My definition enables the calculation of the equilibrium conditions
and entropy changes.
The additive constant, $X$, may be determined by convention.

\section{Peters'  objection}
\label{section: Peters}

A prominent  question  
in   the literature is
whether 
entropy should be defined in one step or two.
The two-step approach 
can be described as hybrid
because it
starts with a definition of entropy,
notes that the definition 
fails in some respect,
and then corrects it 
to agree 
more closely 
with the thermodynamic
properties of entropy.
The historical  reason for this peculiar question
lies in the effort
 to maintain
a  definition of entropy 
in the form 
of the logarithm of a volume in phase space
by modifying it to correct 
the dependence on particle 
number\cite{Dieks_unique_entropy,Dieks_logic_of_identity_2014,Peters_2010,DV,VD,Cheng}.
Since this process usually involves the inclusion 
of a negative term,
$-k_B \ln N!$,
the result is often  called a 
``reduced entropy.''

Peters has 
introduced 
an interesting  hybrid definition of the entropy\cite{Peters_2010,Peters_2014}.
In doing so, he explicitly rejected 
the derivation of entropy given in 
Section \ref{section: statistical mechanics},  
although his only criticism 
turns out to be something we agree on.
We both
 recognized that 
macroscopic experiments  
do not identify individual particles,
so we can never know which 
particles are in which system.
However,
Peters claimed that my version was
``imprecise''
because it did not include the condition 
he denoted as being ``harmonic,''
defined as follows.
\begin{quote}
Systems for
which all possible particle compositions are equiprobable 
will be called harmonic\cite{Peters_2010}.
\end{quote}
For comparison,
I had written  that,
\begin{quote}
when a system of
distinguishable particles is allowed to exchange particles
with the rest of the world, we must include the permutations
of all possible combinations of particles that might enter or
leave the system\cite{RHS_4}.
\end{quote}
It is clear that
we have made   
essentially the same assumption.

Peters'  takes a
 hybrid approach
in that  he chooses 
to define 
a form of the Shannon entropy,
and then ``reduces'' it to arrive at 
the final form\cite{Shannon,Peters_2010}.
\begin{eqnarray}
R_P
&=&
-k_B
\sum_{i=1}^M
\int d^{3N_i}p_i \, \int d^{3N_i}q_i  \nonumber \\
&&
\times \rho_i(p_i,q_i)
 \ln \left( \rho_i(p_i,q_i) h^{3N_i}   \right)  \nonumber \\
&&
 -k_B \ln N!
\end{eqnarray}
This form does have the correct $N$-dependence,
and for the correct reason.
However,
$R_P$
  fails to  
satisfy the second law of thermodynamics.

In Section 4.3.3.2 of 
Ref.~\cite{Peters_2010},
Peters discusses an irreversible process 
initiated by the release 
of  constraints
to allow exchange of energy and particles 
between two subsystems.
He  assumes that 
``both before and after the exchange''
 the two subsystems  
``are in  microcanonical equilibrium.''
The problem is that 
this assumption
is contradicted by 
Liouville's theorem,
which  
requires the total time derivative 
of the probability distribution in the phase space 
of the complete composite system to vanish.
This  means
$R_P$  does not increase 
during an irreversible process,
so it does not satisfy 
the second law of thermodynamics.

Peters  explicitly 
acknowledges the difficulty posed by 
Liouville's theorem
in his Section 5.6.5,
writing 
that, 
``the Liouville equation is entropy
conserving  and therefore cannot describe irreversible processes.''
He does not comment on the  contradiction 
between  his
 Sections 4.3.3.2 and 5.6.5.

In contrast,
the Liouville equation 
does not conserve the entropy 
as defined in this paper,
and the Second Law is satisfied.

\section{Summary}
\label{section: summary}

I've argued for a definition of the thermodynamic entropy
based on 
the probability distribution of 
the macroscopic variables in a composite system.
The entropy defined this way satisfies 
 the postulates
for thermodynamics\cite{Tisza,Callen,RHS_book}.
I've 
addressed 
the objections 
by Dieks\cite{Dieks_unique_entropy,Dieks_logic_of_identity_2014}
 and Peters\cite{Peters_2010,Peters_2014}
to this  derivation of the entropy
from statistical mechanics
and shown that 
they are not valid.

 Since the thermodynamic entropy 
 is known to be unique
 apart from constants  chosen by convention\cite{Lieb_Yngvason},
  any other valid   definition of the entropy 
  must 
  be equivalent the one presented here.

\section*{Acknowledgement}

I would like to thank 
Roberta Klatzky 
for many helpful discussions.
This research did not receive any specific grant from funding agencies in the public, commercial, or
not-for-profit sectors.

\makeatletter
\renewcommand\@biblabel[1]{#1. }
\makeatother

\bibliography{Swendsen_ENTROPY}

\begin{thebibliography}{31}
\bibitem [{foo(){\natexlab{a}}}]{footnote_1_particle_types}%
  \BibitemOpen
  \bibinfo {note} {I will only write the equations for a single type of
  particle. The generalization to several different kinds of particles is
  trivial, but requires indices that might obscure the essential
  argument.}\BibitemShut {Stop}%
\bibitem [{foo(){\natexlab{b}}}]{footnote_on_Dieks_notation}%
  \BibitemOpen
  \bibinfo {note} {Readers interested in following the original phrasing of the
  argument in Dieks' paper may encounter some difficulties caused by the
  notation\cite{Dieks_unique_entropy}. The paper begins by using the symbol $N$
  to denote the number of particles in a simple system. When a composite system
  is first discussed in connection with Dieks' Eq.(1), the particle numbers for
  the two simple subsystems are $N_1$ and $N_2$, while $N=N_1+N_2$ has become a
  constant that denotes the total number of particles in the isolated composite
  system. In his Eq.(3), $N$ is back to being a variable that denotes the
  particle number in a simple system. In his Eq.(4), $N$ is again the total
  number of particles in the isolated composite system, and again a
  constant.}\BibitemShut {Stop}%
\end{thebibliography}

\end{document}